\documentclass[a4paper,10pt]{article}

\usepackage{amsmath}

\usepackage[english]{babel} % the language of the conference
\usepackage{graphicx} % used to insert the figure
\usepackage{multirow} % used for the table
\usepackage[center]{caption}
\usepackage{geometry}
\usepackage{indentfirst}
\usepackage{txfonts}

\graphicspath{{Figures_artISMA2014/}}

\newcommand{\EV}[1]{\mathbb{E}\left[#1\right]}

\usepackage{cases}
\usepackage{subfigure}
\usepackage[table]{xcolor}
\usepackage[utf8x]{inputenc}
\usepackage[T1]{fontenc}
\usepackage{color}
\usepackage{abstract}

\usepackage{hyperref}
\hypersetup{
    colorlinks,%
    citecolor=red,%
    filecolor=black,%
    linkcolor=blue,%
    urlcolor=black
}

\title{\bfseries Prediction of the dynamic oscillation threshold of a clarinet model: Comparison between analytical predictions and simulation results}

\author{B. Bergeot$^{a,b,*}$ , A. Almeida$^{b,c}$, C. Vergez$^{a}$ and B. Gazengel$^{b}$\vspace{12pt}\\
{\small$^{a}$LMA, CNRS UPR7051, Aix-Marseille Univ., Centrale Marseille, F-13402 Marseille Cedex 20, France}\\
 {\footnotesize$^{b}$LUNAM Universit\'{e}, Universit\'{e} du Maine, UMR CNRS 6613, Laboratoire d’Acoustique, Avenue Olivier Messiaen, 72085 Le Mans Cedex 9, France}\\
{\footnotesize$^{c}$ School of Physics, The University of New South Wales, Sydney UNSW 2052, Australia}
}

\date{\itshape Proceeding of the Stockholm Music Acoustics Conference (SMAC)\\
30 July — 3 August 2013\\
KTH Royal Institute of Technology, Stockholm, Sweden}

\begin{document}

\twocolumn[
\begin{@twocolumnfalse}

\maketitle

\begin{abstract}
\noindent Simple models of clarinet instruments based on iterated maps have been used in the past to successfully estimate the threshold of oscillation of this instrument as a function of a constant blowing pressure. However, when the blowing pressure gradually increases through time, the oscillations appear at a much higher value than what is predicted in the static case. This is known as bifurcation delay, a phenomenon studied in~\cite{BergeotNLD2012} for a clarinet model. In numerical simulations the bifurcation delay showed a strong sensitivity to numerical precision.

% Equations modeling the clarinet are known to be reducible to a nonlinear iterated map. In this simple model, when the blowing pressure varies, a phenomenon of bifurcation delay appears. When the model is simulated, it is shown in~\cite{BergeotNLD2012} that the bifurcation delay is very sensitive to the numerical precision. 

\noindent This paper presents an analytical estimation of the bifurcation delay of the simplified clarinet model taking into account the numerical precision of the computer. The model is then shown to correctly predict the bifurcation delay in numerical simulations.
\end{abstract}
\vspace{24pt}
\end{@twocolumnfalse}]

\section{Introduction}\label{sec:introduction}

The oscillation threshold of the clarinet has been extensively studied in the literature \cite{KergoActa2000,dalmont:3294} assuming that the blowing pressure is constant. In this context, the \textit{static} oscillation threshold $\gamma_{st}$ is defined as the minimum value of the blowing pressure for which there is a periodic oscillating regime. This value of the threshold is obtained by applying a constant blowing pressure, allowing enough time to let the system reach a permanent regime (either static or strictly periodic), and repeating the procedure for other constant blowing pressures. Most studies using iterated maps  are restricted to static cases, even if transients are observed. They focus on the asymptotic amplitude regardless of the history of the system and of the history of the control parameter.

 A recent article~\cite{BergeotNLD2012} studied the behavior of a clarinet model when the blowing pressure increases linearly. The model starts its oscillations for a much higher value of the blowing pressure than the  \textit{static} oscillation threshold. An analytical expression of this dynamical threshold has been derived and its properties studied: the dynamic threshold does not depend on the increase rate of the blowing pressure (for sufficiently low increase rates), but  is very sensitive to the value of the blowing pressure at which the increase is started.

The article~\cite{BergeotNLD2012} ends with a comparison between the analytical predictions and numerical simulations (Fig.~10 in~\cite{BergeotNLD2012}), revealing an important sensitivity to the precision used in numerical simulations. Indeed, numerical results only converge towards theoretical ones when the model is computed with hundreds or thousands of digits (i.e. approximating an  infinitely precise simulation). Otherwise, the observations of numerically simulated thresholds are far from the theoretical ones, they depend on the increase rate of the blowing pressure and are independent of the starting value of the blowing pressure. The conclusion is that theoretical results obtained in~\cite{BergeotNLD2012} cannot explain the behavior of the model simulated in the common double-precision of a modern CPU.

The aim of this paper is to explain and predict the start of the oscillations in simulations performed with the usual double-precision of computer CPUs (around 15 decimal debits). A model introduced in a journal article \cite{BergeotNLD2012} is modified to predict the oscillation threshold in limited precisions. The expression of this oscillation threshold is given in section \ref{sec:3}. In the same section,  the theoretical thresholds (static and dynamic ignoring or taking into account the precision) are compared to the thresholds observed in numerical simulations. The influence of the speed at which the blowing pressure is increased is discussed, as well as that of the initial value of the blowing pressure. A similar analysis is made for the second control parameter of the model (related to opening of the embouchure at rest). The clarinet model and major results  from~\cite{BergeotNLD2012} are first briefly recalled in section \ref{sec:STA}.

\section{State of the art}\label{sec:STA}

\subsection{Clarinet Model}\label{sec:ClarMod}

% * Their behavior in steady state is well-known (from a small set of parameters it is possible to determine whether the instrument state is steady, oscillating or chaotic

% * However this knowledge is mostly asymptotic, we have the tools to describe the steady state, but not many works have been dedicated to understanding the transient behavior

% Model
% How do we describe the instrument mathematically?
% Two elements: exciter + resonator
This model divides the instrument into two elements: the exciter and the resonator.
The exciter is  modeled by a nonlinear function $F$ also called nonlinear characteristic of the exciter, which relates the pressure applied to the reed $p(t)$ to the flow $u(t)$ through its opening. The resonator (the bore of the instrument) is described by its reflection function $r(t)$. $p$ and $u$ are two non-dimensional state variables that are sufficient to describe the state of the instrument.

% The parameters
The solutions $p(t)$ and $u(t)$ depend on the control parameters: $\gamma$ proportional to the mouth pressure $P_m$ according to
\begin{equation}
  \label{eq:1}
  \gamma = \frac{P_m}{P_M} = \frac{P_m}{kH}
\end{equation}
where $P_M$ represents the pressure needed to close the reed entrance (also used to normalize the pressure $p(t)$) a product of $1/k$ the acoustic compliance of the reed and $H$ its distance to the lay at rest. The other parameter is $\zeta$ which is related to the opening of the embouchure at rest according to the formula
\begin{equation}
  \label{eq:2}
  \zeta ={Z_c \,U_A}/{P_M}=Z_c w H \sqrt{\frac{2}{\rho P_M}} .
\end{equation}
Here, $Z_c$ is the characteristic impedance at the input of the bore, $w$ the effective width of the reed, and $U_A$ the maximum flow admitted by the reed valve. For most of the analysis below, this parameter is fixed at $0.5$, a typical value observed in musicians, but the analysis can easily be reproduced for other values of $\zeta$. The nonlinear characteristic is provided by the Bernoulli equation describing the flow in the reed channel \cite{HirschAal1990,MOMIchap7}. 

% A method to understand the behavior of the instrument: iterated maps
The model is extremely simplified by considering a straight resonator in which the eventual losses are independent of frequency. In the current work, losses are neglected in all calculations. The reed is considered as an ideal spring \cite{Maga1986,MechOfMusInst,KergoCFA2004,OllivActAc2005,dalmont:3294,Cha08Belin}. With these assumptions, the reflection function becomes a simple delay with sign inversion. Using the variables $p^+$ and $p^-$ (outgoing and incoming pressure waves respectively) instead of the variables $p$ and $u$, the system can be simply described by an iterated map \cite{Maga1986}:

\begin{equation}
p^+_n=G\left(p^+_{n-1},\gamma \right).
\label{Def_stat}
\end{equation}

An explicit expression for this function is given by Taillard \cite{NonLin_Tail_2010} for $\zeta<1$. This function depends on the control parameters $\gamma$ and $\zeta$. The time step $n$ corresponds to the round trip time $\tau=2l/c$ of the wave with velocity $c$ along the resonator of length $l$.

%\begin{figure}[t]
%\centering
%\includegraphics[width=0.84\columnwidth]{ArtSMAC_fig0}
%\caption{Iteration function $G$ for $\gamma=0.42$ and $\zeta=0.6$.}
%\label{fig11}
%\end{figure}

\begin{figure}[t]
\centering
\includegraphics[width=0.86\columnwidth]{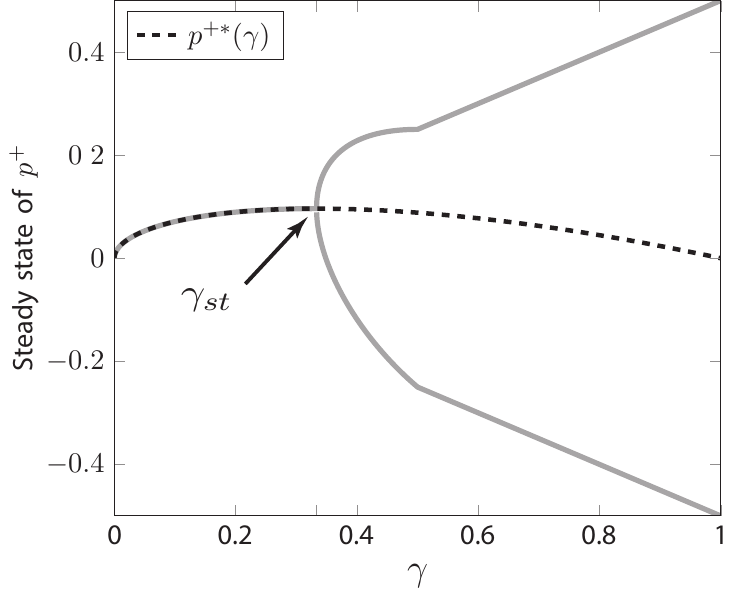}
\caption{Graphical representation of the static bifurcation diagram for $\zeta=0.5$. In gray, the stable solutions, in dashed black, the non-oscillating solution.}
\label{fig12}
\end{figure}

Using the universal properties of the iterated maps \cite{Feigen1978,Feigen1979},  useful information about the instrument behavior can be drawn from the study of the iteration function. So far, these studies come from the \textit{static} bifurcation theory, which assumes that the control parameter $\gamma$ is constant. For instance, it is possible to determine the steady state of the system as a function of the parameter $\gamma$, and to plot a bifurcation diagram shown in figure \ref{fig12}. When no losses are considered, the oscillation threshold  $\gamma_{st}$ is:

\begin{equation}
\gamma_{st} =\frac{1}{3},
\end{equation}
For all values of the control parameter $\gamma$ below $\gamma_{st}$ the series $p^+_n$ converges to a single value $p^{+*}$ corresponding to the fixed point of the function $G$, i.e.~the solution of  $p^{+*}=G_{\gamma}\left(p^{+*}\right)$. When the control parameter $\gamma$ exceeds $\gamma_{st}$ the fixed point of $G$ becomes unstable and the steady state becomes a 2-valued oscillating regime. Figure \ref{fig12} shows an example of the bifurcation diagram with respect to the variable $p^+$.

An iterated map approach can be used to predict the asymptotic (or \textit{static}) behavior of an ideal clarinet as a function of a constant mouth pressure. This procedure avoids the phenomenon of \textit{bifurcation delay} which is observed in numerical simulations when the control parameter $\gamma$ is increased.

\subsection{Slowly time-varying mouth pressure}\label{sec:2:2}

\subsubsection{Dynamic bifurcation}

A control parameter $\gamma$ increasing linearly with time is taken into account by replacing eq.~(\ref{Def_stat}) by eqs.~(\ref{dynsys_pp_a}) and (\ref{dynsys_pp_b}):

\begin{subnumcases}{\label{dynsys_pp}}
p^+_n=G\left(p^+_{n-1},\gamma_n\right)\label{dynsys_pp_a}\\
\gamma_{n}=\gamma _{n-1} +\epsilon.\label{dynsys_pp_b}
\end{subnumcases}

\begin{figure*}[t]
\centering
\subfigure[precision = 100]{\includegraphics[width=1.04\columnwidth]{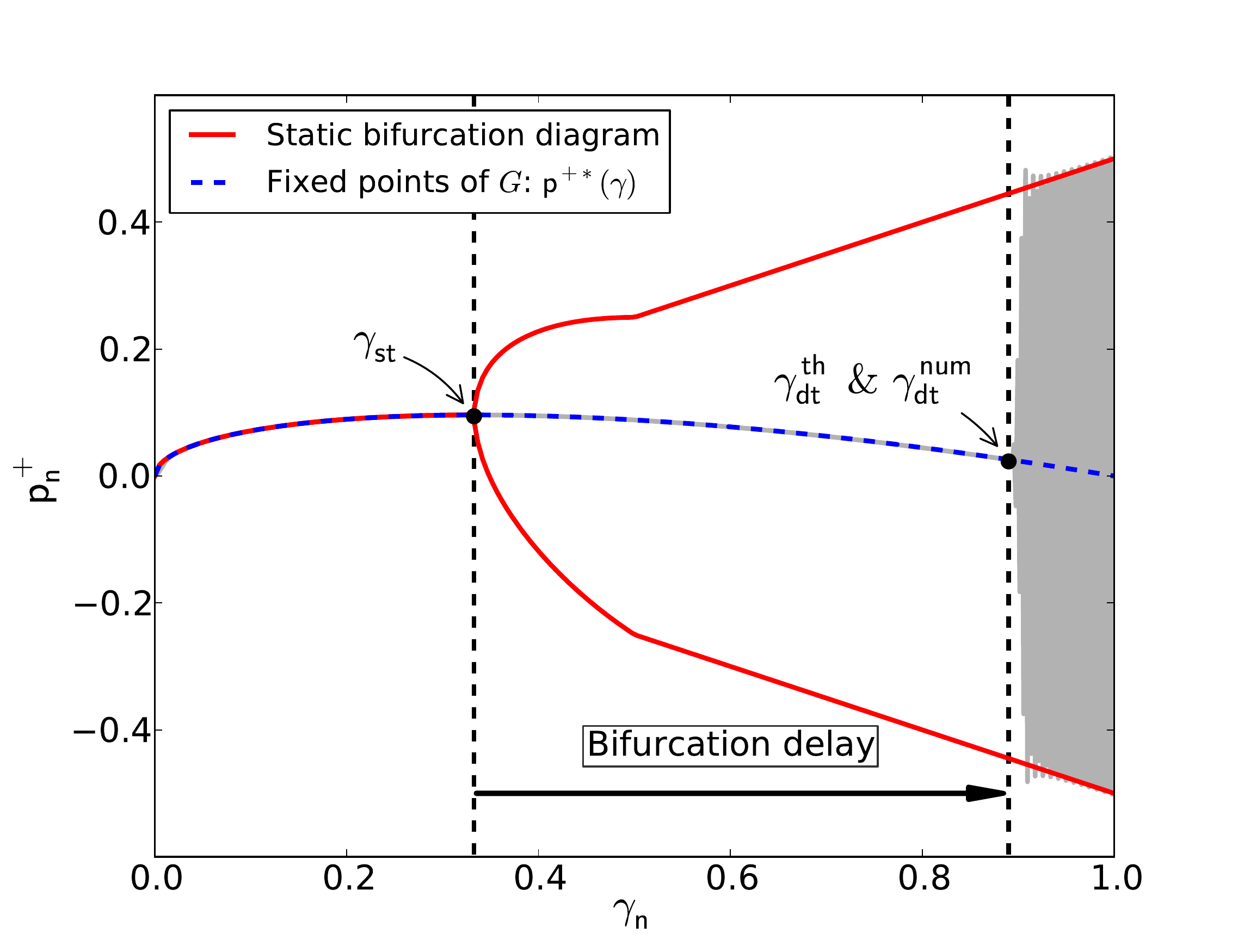}\label{fig:2a}}
\subfigure[precision =15]{\includegraphics[width=1.04\columnwidth]{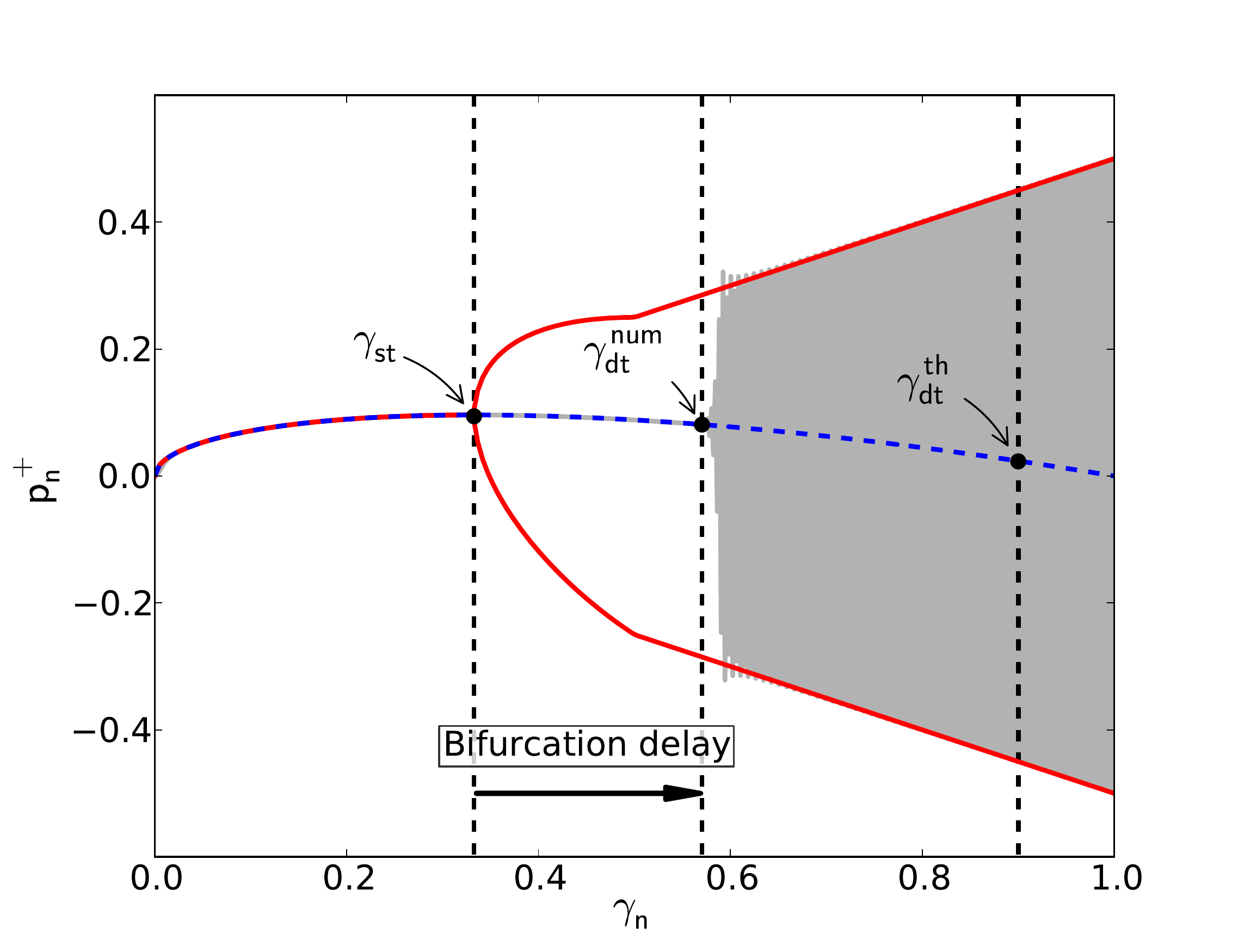}\label{fig:2b}}
\caption{Comparison between \textit{static} and \textit{dynamic} bifurcation diagram as functions of $\gamma_n$. $\epsilon=2\cdot10^{-3}$, $\zeta=0.5$ and the numerical precision is equal to 100 (figure \ref{fig:2a}) and 15 (figure \ref{fig:2b}) decimal digits. The thresholds $\gamma_{st}$, $\gamma_{dt}^{th}$ and $\gamma_{dt}^{num}$ are represented.}
\label{fig:2}
\end{figure*}

The parameter $\gamma$ is assumed to increase slowly, hence $\epsilon$ is considered arbitrarily small ($\epsilon\ll 1$). When the series $p^+_n$ is plotted with respect to parameter $\gamma_n$ the resulting curve can be interpreted as a \textit{dynamic} bifurcation diagram and it can be compared to the \textit{static} bifurcation diagram (fig.~\ref{fig:2}).

Because of the time variation of $\gamma$, the system (\ref{dynsys_pp}) is subject to the phenomenon of bifurcation delay \cite{Baesens1991,Fruchard2007}: the bifurcation point is shifted from the \textit{static oscillation threshold} $\gamma_{st}$~\cite{dalmont:3294} to the \textit{dynamic oscillation threshold} $\gamma_{dt}$~\cite{BergeotNLD2012}. The difference $\gamma_{dt}-\gamma_{st}$ is called the \textit{bifurcation delay}. 

The techniques used in dynamic bifurcation theory are now required to properly analyze the system. Article~\cite{BergeotNLD2012} provides an analytical study of the dynamic flip bifurcation of the clarinet model (i.e.~system (\ref{dynsys_pp})) based on a generic method given by Baesens~\cite{Baesens1991}. The main results of this study, leading to a theoretical estimation of the dynamic oscillation threshold of the clarinet are recalled below.

%The time evolution of the series $p^+_n$ in dynamic case is divided into three parts. The series $p^+_n$  first shows a short oscillating transient, which is called \textit{transient oscillating} \textit{dynamic regime}. This oscillation decays into a \textit{non-oscillating dynamic regime}. Beyond a certain threshold, a new oscillation grows, giving rise to the \textit{final oscillating dynamic regime}. 

%The dynamic oscillation threshold is defined as the value of the parameter $\gamma$ for which the transition (i.e. the bifurcation) from the \textit{non-oscillating dynamic regime} to the \textit{final oscillating dynamic regime} occurs.

\subsubsection{Dynamic oscillation threshold of the clarinet model without noise}
\label{sec:dynoscthredet}

A possible theoretical estimation of the dynamic oscillation threshold consists in identifying the value of $\gamma$ for which the orbit of the series $p^+_n$ escapes from a neighborhood of arbitrary distance of an \textit{invariant curve} $\phi(\gamma,\epsilon)$. More precisely, the dynamic oscillation threshold is reached when the distance between the orbit and the invariant curve becomes equal to $\epsilon$.

The invariant curve (i.e. invariant under the mapping (\ref{dynsys_pp}), described for example in \cite{FruchaScaf2003}) can be seen as the equivalent of a fixed point in static regimes, functioning as an attractor for the state of the system. It satisfies the following equation: 
\begin{equation}
\phi(\gamma,\epsilon)=G\left(\phi(\gamma-\epsilon,\epsilon),\gamma\right).
\label{eqdiff_1}
\end{equation}
The procedure to obtain the theoretical estimation $\gamma_{dt}^{th}$ of the dynamic oscillation threshold is as follows: a theoretical expression of the invariant curve is found for a particular (small) value of the increase rate $\epsilon$ (i.e. $\epsilon \ll1$). The system~(\ref{dynsys_pp}) is then expanded into a first-order Taylor series around the invariant curve and the resulting linear system is solved analytically. Finally, $\gamma_{dt}^{th}$ is derived from the analytic expression of the orbit.

The analytic estimation of the dynamic oscillation threshold $\gamma_{dt}^{th}$ is defined in~\cite{BergeotNLD2012}:
\begin{equation}
\int_{\gamma_0+\epsilon}^{\gamma_{dt}^{th}+\epsilon}\ln\left| \partial_xG\left(\phi(\gamma'-\epsilon),\gamma'\right)\right|d\gamma'=0,
\label{dynoscthre_2}
\end{equation}
where $\gamma_0$ is the initial value of $\gamma$ (i.e. the starting value of the linear ramp). The main properties of $\gamma_{dt}^{th}$ are (Fig.~6 of ~\cite{BergeotNLD2012}):
\begin{itemize}
\item $\gamma_{dt}^{th}$ does not depend on the slope of the ramp $\epsilon$ (provided $\epsilon$ is small enough) 
\item $\gamma_{dt}^{th}$  depends on the initial value $\gamma_0$ of the ramp.
\end{itemize}

\section{Numerical simulations: the precision cannot be ignored}\label{sec:3}

\subsection{Problem statement}

The above theoretical prediction ignores the round-off errors of the computer. The bifurcation delays $\gamma_{dt}^{num}$ observed in simulations\footnote{In simulations, $\gamma_{dt}^{num}$ is estimated as the value for which the distance between the simulated orbit and the invariant curve becomes equal to $\epsilon$.} are seen to converge to the theoretical ones for very high numerical precision, typically when hundreds or thousands digits are considered in the simulation (cf.~figure \ref{fig:2a} where a precision\footnote{The choice of the precision is possible using \emph{mpmath}, the arbitrary precision library of \emph{Python}.} of 100 was used). However, in standard double-precision arithmetic (precision close to 15 decimals), theoretical predictions of the dynamic bifurcation point $\gamma_{dt}^{th}$ are far from the thresholds $\gamma_{dt}^{num}$ observed in numerical simulations. An example is shown in figure \ref{fig:2b}. In particular, the numerical bifurcation threshold depends on the slope $\epsilon$, unlike the theoretical predictions $\gamma_{dt}^{th}$. Moreover, the dependence of  the bifurcation point on the initial value $\gamma_0$ is lost over a wide range of $\gamma_0$.

As a conclusion, because they ignore the round-off errors of the computer, theoretical results obtained in~\cite{BergeotNLD2012} fail to predict the behavior of numerical simulations carried out at usual numerical precision. In particular this is problematic when studying the behavior of a synthesis model, or simply when trying to understand the large delays in the threshold of oscillation in simulations of real systems. Following a general method given by Baesens~\cite{Baesens1991}, we show in section~\ref{sec:32} how the dynamic oscillation thresholds of simulations with finite precision can be analytically predicted.

\subsection{Theoretical estimation of the dynamic oscillation threshold in presence of noise}
\label{sec:32}

Following usual modeling of quantization as a uniformly distributed random variable \cite{oppenheim1989discrete}, round-off errors of the computer are introduced as $\xi_n$ (referred as an additive white noise). Therefore, system (\ref{dynsys_pp}) becomes:

\begin{subnumcases}{\label{dynsys_pp_no}}
p^+_n=G\left(p^+_{n-1},\gamma_n\right)+\xi_n \label{dynsys_pp_no_a}\\
\gamma_{n}=\gamma _{n-1} +\epsilon,
\end{subnumcases}
where $\xi_n$ is a white noise with an expected value equal to zero (i.e. $\EV{\xi_n}=0$) and variance $\sigma$ defined by:
\begin{equation}
\EV{\xi_m \xi_n} \, =\sigma^2\delta_{mn},
\label{eq:noiselev}
\end{equation}
where $\delta_{mn}$ is the Kronecker delta. The definition of the expected value $\mathbb{E}$ is provided in \cite{IntroProbShel}. Equations (\ref{dynsys_pp_no}) are used for the analytic study. In later sections, the results of this analytical study will be compared to numerical simulations of the system~(\ref{dynsys_pp}) using a numerical precision of 15 decimals. As a consequence, the noise level $\sigma$ will be equal to $10^{-15}$.
 
The method to obtain the theoretical estimation of the dynamic oscillation threshold which take into account the precision (noted $\hat{\gamma}_{dt}^{th}$) is the same as to obtain $\gamma_{dt}^{th}$ (cf. section \ref{sec:dynoscthredet}). In addition, because of the noise the bifurcation delay is reduced so that the dynamic oscillation threshold $\gamma_{dt}$ is assumed to be close\footnote{This hypothesis could be questioned because according to figures~\ref{fig:2a}, even in the presence of noise, the bifurcation delay can be large. However, this hypothesis is required to carry out calculations.} to the static oscillation threshold $\gamma_{st}$. Using this approximation, the expression of $\hat{\gamma}_{dt}^{th}$ is:

\begin{equation}
\hat{\gamma}_{dt}^{th} = \gamma_{st}+\sqrt{-\frac{2\epsilon}{K}\ln\left[\left(\frac{\pi}{K}\right)^{1/4}
\frac{\sigma}{\epsilon^{5/4}}\right]},
\label{eq:seuidynno}
\end{equation}
which is the theoretical estimation of the dynamic oscillation threshold of the stochastic systems (\ref{dynsys_pp_no}) (or of the system (\ref{dynsys_pp}) when it is computed using a finite precision). $K$ is a constant that depends on the slope of $\partial_x G(p^+(\gamma),\gamma)$, the derivative of the iteration function at the fixed point. 

A summary table of different notations of the oscillation thresholds is provided in table \ref{ta:TaOfNotThresh}.

\begin{table}[h]
\centering
{%\small
\rowcolors{1}{white}{gray!25}
\begin{tabular}{|p{0.8cm}|p{5.8cm}|}
\hline
\multicolumn{2}{ |c| }{\textbf{Table of Notation}}\\ \hline
$\gamma_{st}$ & static oscillation threshold \\
%$\gamma_{dt}$ & dynamic oscillation threshold \\
$\gamma_{dt}^{th}$ & theoretical estimation of the dynamic oscillation threshold of the clarinet model without noise\\
$\hat{\gamma}_{dt}^{th}$ & theoretical estimation of the dynamic oscillation threshold in presence of noise  \\
$\gamma_{dt}^{num}$  & dynamic oscillation threshold calculated on numerical simulations \\
\hline
\end{tabular}
}
\caption{Table of notation for thresholds of oscillation.}
\label{ta:TaOfNotThresh}
\end{table}

\subsection{Benchmark of theoretical estimators for the dynamic oscillation threshold}
\label{sec:Bench}

This section compares the  theoretical estimation of the dynamic oscillation threshold $\hat{\gamma}_{dt}^{th}$ for a standard deviation $\sigma=10^{-15}$ with the thresholds observed in numerical simulations using the regular 64-bit double-precision of a CPU (about 15 decimal digits). The comparison is carried out as a function of the increase rate ($\epsilon$) of the blowing pressure,  the initial value $\gamma_0$ and the embouchure parameter $\zeta$. The estimations of the theory with noise ($\hat{\gamma}_{dt}^{th}$) are plotted simultaneously with $\gamma_{st}$ and $\gamma_{dt}^{th}$.

\begin{figure}[h]
\centering
\includegraphics[width=1\columnwidth]{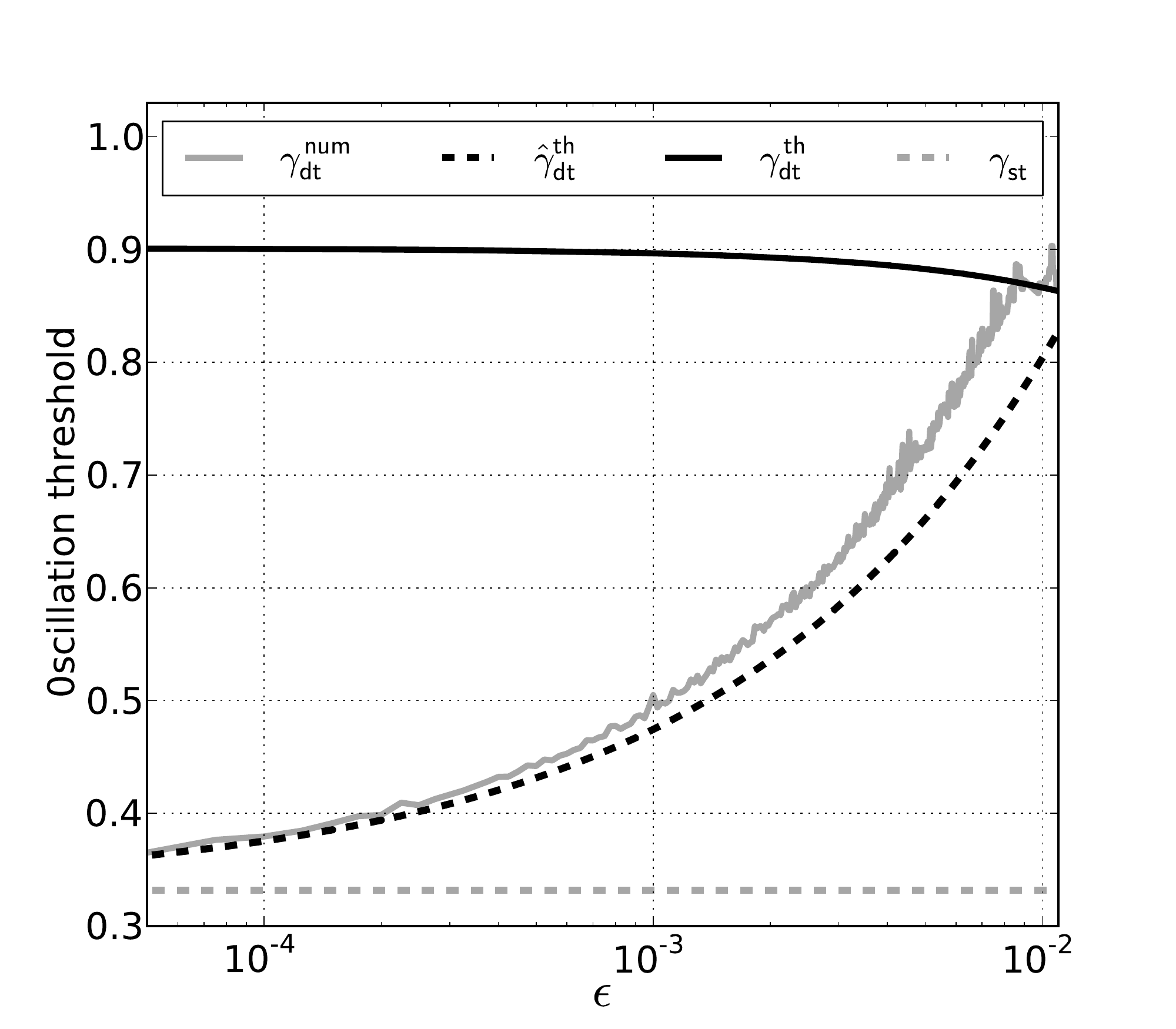}
\caption{Graphical representation of $\gamma_{dt}^{num}$ with respect to the slope $\epsilon$, for $\gamma_0 = 0$ and $\zeta=0.5$. Results are compared to analytic \textit{static} and \textit{dynamic} thresholds: $\gamma_{st}$, $\gamma_{dt}^{th}$ and $\hat{\gamma}_{dt}^{th}$.}
\label{fig:InfEps}
\end{figure}

\begin{figure}[h!]
\centering
\includegraphics[width=1\columnwidth]{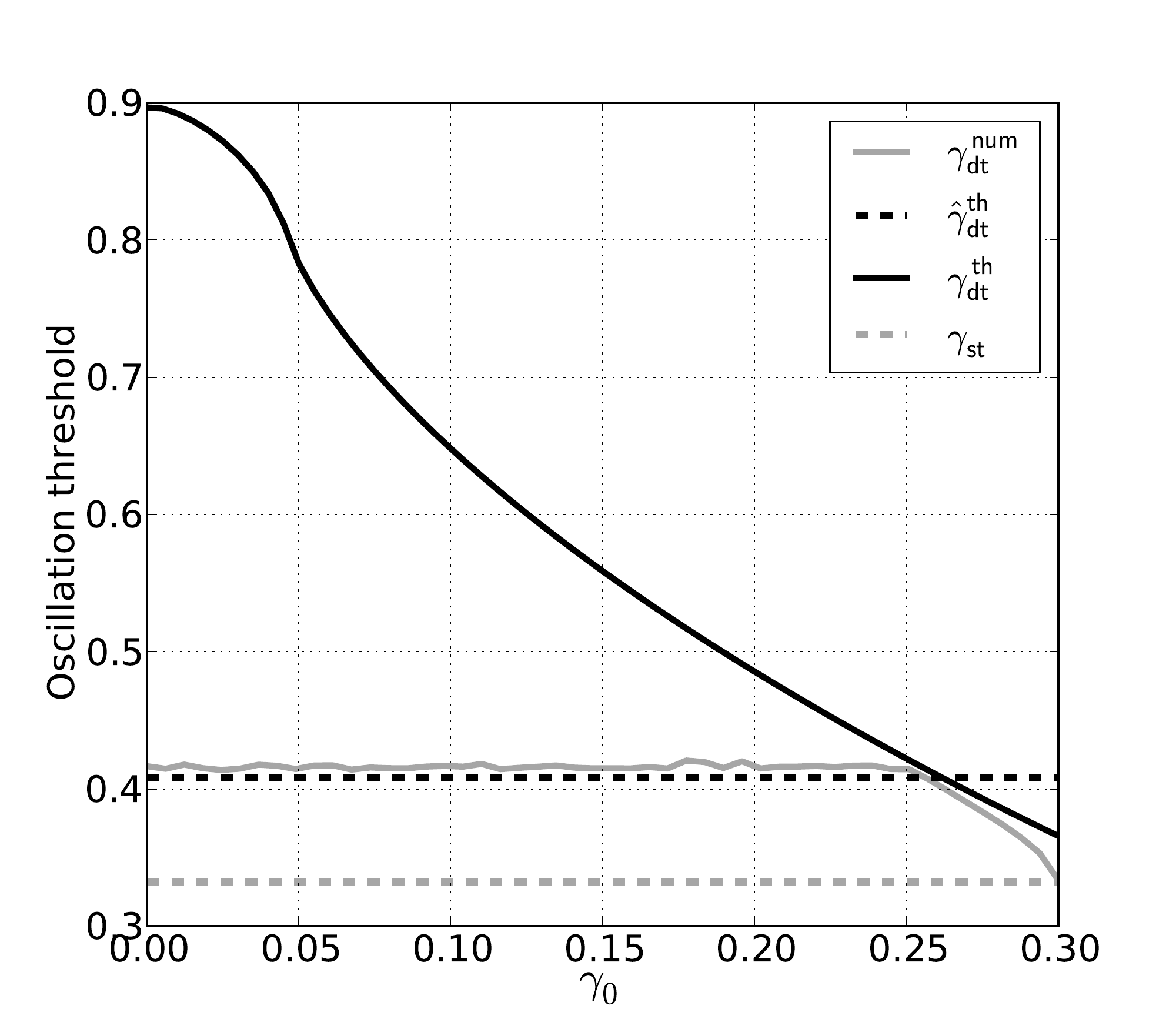}
\caption{Comparison between theoretical prediction of oscillation thresholds (dynamic without noise: $\gamma_{dt}^{th}$ and with noise: $\hat{\gamma}_{dt}^{th}$, and static $\gamma_{st}$) and the dynamic threshold $\gamma_{dt}^{num}$. Various thresholds are plotted with respect to the initial condition $\gamma_0$ with $\epsilon = 3\cdot 10^{-4}$ and $\zeta=0.5$.}
\label{fig:InfGam0}
\end{figure}

\begin{figure*}[t]
\centering
\subfigure[$\epsilon = 10^{-4}$ and $10^{-3}$]{\includegraphics[width=1.02\columnwidth]{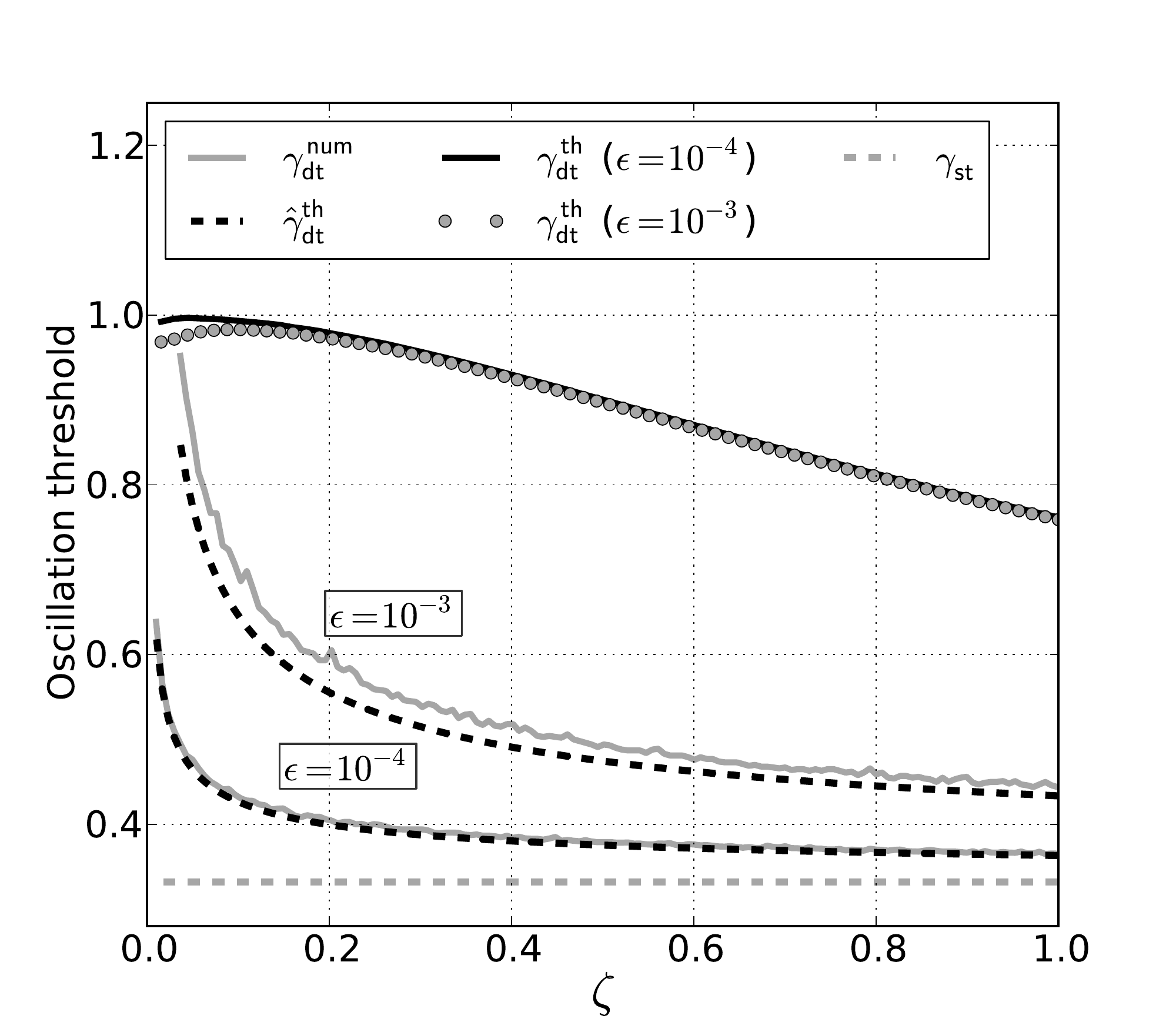}\label{fig:InfZetaa}}
\subfigure[$\epsilon = 10^{-2}$]{\includegraphics[width=1.02\columnwidth]{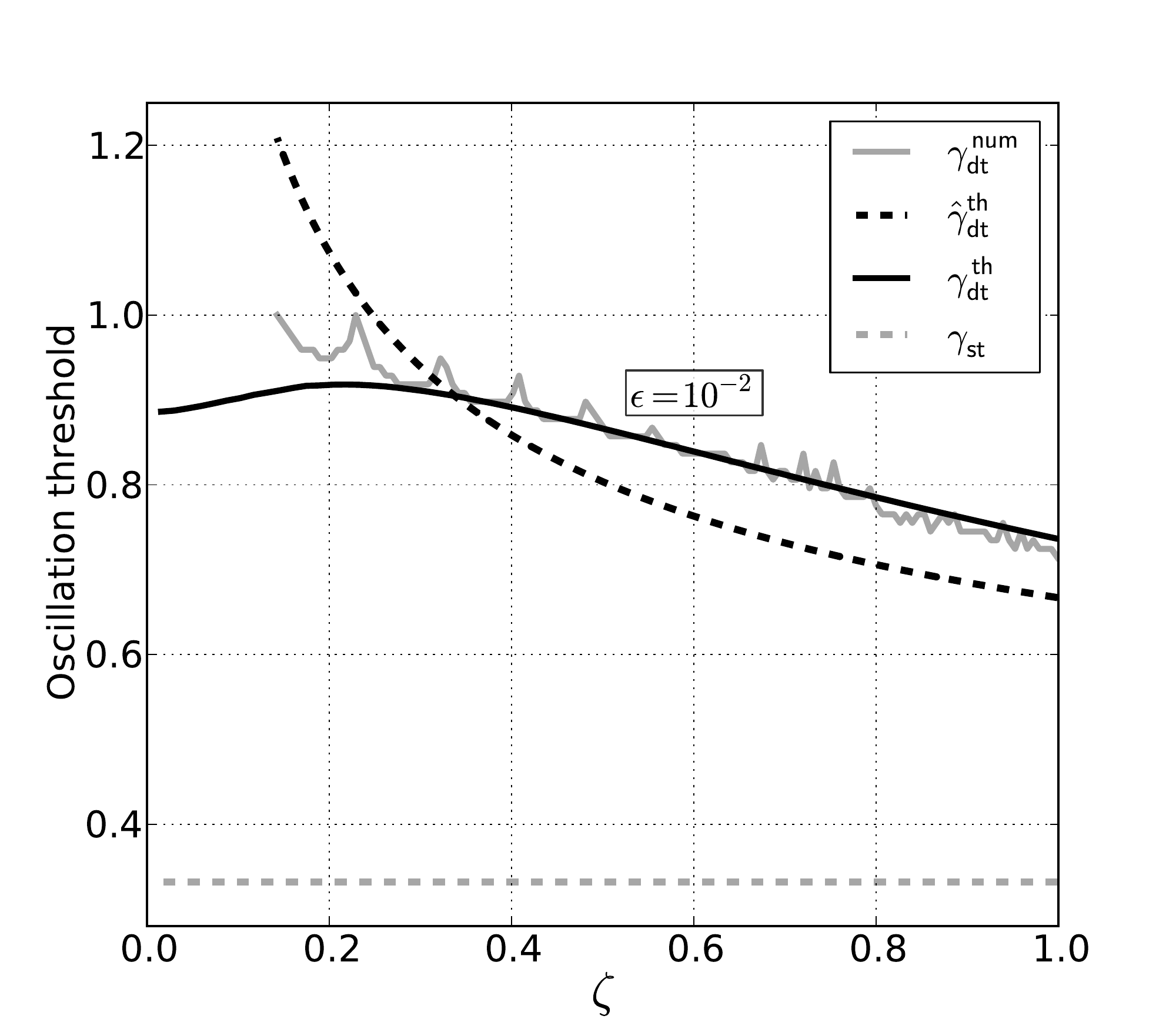}\label{fig:InfZetab}}
\caption{Comparison between theoretical prediction of oscillation thresholds (dynamic without noise: $\gamma_{dt}^{th}$ and with noise: $\hat{\gamma}_{dt}^{th}$, and static $\gamma_{st}$) and the dynamic threshold $\gamma_{dt}^{num}$. Various thresholds are plotted with respect to the embouchure parameter $\zeta$ with $\gamma_0=0$ and (a) $\epsilon = 10^{-4}$ and $10^{-3}$ and (b) $\epsilon = 10^{-2}$.}
\label{fig:InfZeta}
\end{figure*}

In figures \ref{fig:InfEps} and \ref{fig:InfGam0}, the various thresholds are plotted with respect to $\epsilon$ and to $\gamma_0$  respectively. Unlike $\gamma_{dt}^{th}$ (or $\gamma_{dt}^{num}$ calculated on simulations with very high precision, see~\cite{BergeotNLD2012}), here the main properties of $\gamma_{dt}^{num}$ are:

\begin{itemize}
\item $\gamma_{dt}^{num}$ depends on the slope $\epsilon$
\item $\gamma_{dt}^{num}$ does not depend on $\gamma_0$ over a wide range of $\gamma_0$.
\end{itemize}

In both figures \ref{fig:InfEps} and \ref{fig:InfGam0} we observe a good agreement between $\hat{\gamma}_{dt}^{th}$ and $\gamma_{dt}^{num}$. However, for large $\epsilon$, in figure \ref{fig:InfEps}, and for $\gamma_0$ close to the static threshold $\gamma_{st}$, in figure \ref{fig:InfGam0}, the theoretical threshold $\gamma_{dt}^{th}$ for infinite precision is a better prediction of the dynamic threshold. Therefore, in this case, the round-off errors of the computer can be ignored.

In figure \ref{fig:InfZeta}, thresholds are plotted with respect to the embouchure parameter $\zeta$, showing that $\gamma_{dt}^{num}$ decreases with $\zeta$. In figure \ref{fig:InfZetaa}, two increase rates $\epsilon$ are used ($10^{-4}$ and $10^{-3}$). These  slopes are sufficiently small so that the curves for $\gamma_{dt}^{th}$ overlap\footnote{cf.~properties of $\gamma_{dt}^{th}$ in section \ref{sec:dynoscthredet}.} (except for small values of $\zeta$). In these situations, the estimation with noise $\hat{\gamma}_{dt}^{th}$ predicts correctly the observed dynamic thresholds $\gamma_{dt}^{num}$ and, as expected, the prediction is better for the slower increase rate $\epsilon$.

The behavior of the system changes for larger $\epsilon$ (cf.~figure \ref{fig:InfZetab} where $\epsilon=10^{-2}$). First of all, for this value of the slope the dependence of $\gamma_{dt}^{th}$ on $\epsilon$ appears. Moreover, as in figure \ref{fig:InfGam0}, beyond the intersection between $\hat{\gamma}_{dt}^{th}$ and ${\gamma}_{dt}^{th}$ the theoretical estimation for infinite precision, $\gamma_{dt}^{th}$, becomes a better prediction of the bifurcation delay.

\section{Conclusions}

In a simplified model of the clarinet, the threshold in mouth pressure above which the oscillations occur can be obtained  using an iterated map approach. This threshold corresponds to 1/3 of the reed beating pressure, but when the mouth pressure is increased with time, the oscillations start at a much higher value than this static threshold. The dynamic threshold calculated with infinite precision is independent on the rate of increase, depending only on the starting value of the mouth pressure. 

Numerical simulations performed using finite precision show very different results in that the dynamic threshold depends on the increase rate and not on the starting value of the mouth pressure. A modified dynamic bifurcation theory including the effect of a stochastic variation in mouth pressure can be derived to correctly approximate the dynamic threshold when the precision is limited.

With a precision of $10^{-15}$, this theory is seen to match the
simulations performed with double-precision, showing that the
threshold is situated between the static threshold $\gamma=1/3$ and
the dynamic threshold (close to $\gamma\simeq 0.9$ for typical playing
conditions). This threshold increases with the rate of increase of
mouth pressure $\epsilon$, and does not depend on the initial
condition $\gamma_0$ throughout most of the range below the static
threshold. Although not shown here, the model can be applied to other
values of precision. Because it is based on a stochastic analysis of
the dynamic system, the model is expected to describe the behavior of
the system subject for instance to turbulence noise.

\section*{acknowledgments}
This work is part of the project SDNS-AIMV ``Syst\`{e}mes Dynamiques Non-Stationnaires - Application aux Instruments \`{a} Vent'', sponsored by \emph{Agence Nationale de la Recherche} (ANR).

%%%%%%%%%%%%%%%%%%%%%%%%%%%%%%%%%%%%%%%%%%%%%%%%%%%%%%%%%%%%%%%%%%%%%%%%%%%%%
%bibliography here

\end{document}